\documentclass[journal]{IEEEtran}
\usepackage{cite}

\usepackage{braket}
\usepackage{array}
\usepackage{adjustbox}
\usepackage{subcaption}
\usepackage{multirow}
\usepackage{booktabs}
\usepackage{amssymb,amsthm}
\usepackage{graphicx}
\usepackage{algorithm}
\usepackage{algorithmic}

\usepackage[caption=false,font=footnotesize]{subfig}

\ifCLASSINFOpdf
\else
\fi
\usepackage{amsmath}
\usepackage{mathtools}
\usepackage{amssymb}
\usepackage{dsfont}
\usepackage[font=footnotesize]{caption}
\usepackage{algorithmic}
	\usepackage{url}
	
	\usepackage{mathtools}
	\usepackage{makecell}

\begin{document}
		%
		\title{Dual Mind World Model Inspired Network Digital Twin for Access Scheduling}
		
		%
		%
		%
		
		\author{Hrishikesh Dutta,~\IEEEmembership{Member,~IEEE,}
			Roberto Minerva,~\IEEEmembership{Senior Member,~IEEE,}
			and~Noel Crespi,~\IEEEmembership{Senior Member,~IEEE}
			\thanks{The authors are with the Data Intelligence and Communication Engineering Lab, Telecom SudParis, Institut Polytechnique de Paris, France (email: hrishikesh.dutta@telecom-sudparis.eu; roberto.minerva@telecom-sudparis.eu; noel.crespi@telecom-sudparis.eu)}
		}
		
		%
		%

	\maketitle
	
	\begin{abstract}
	Emerging networked systems such as industrial IoT and real-time cyber-physical infrastructures demand intelligent scheduling strategies capable of adapting to dynamic traffic, deadlines, and interference constraints. In this work, we present a novel Digital Twin-enabled scheduling framework inspired by Dual Mind World Model (DMWM) architecture, for learning-informed and imagination-driven network control. Unlike conventional rule-based or purely data-driven policies, the proposed DMWM combines short-horizon predictive planning with symbolic model-based rollout, enabling the scheduler to anticipate future network states and adjust transmission decisions accordingly. We implement the framework in a configurable simulation testbed and benchmark its performance against traditional heuristics and reinforcement learning baselines under varied traffic conditions. Our results show that DMWM achieves superior performance in bursty, interference-limited, and deadline-sensitive environments, while maintaining interpretability and sample efficiency. The proposed design bridges the gap between network-level reasoning and low-overhead learning, marking a step toward scalable and adaptive NDT-based network optimization.
	\end{abstract}
	

	%
\IEEEpeerreviewmaketitle

\section{Introduction}
\label{intro}

The next generation of wireless communication systems, such as smart city infrastructure, industrial IoT, and autonomous networks, demands unprecedented levels of adaptability, reliability, and intelligence. These systems must operate under bursty traffic, stringent Quality of Service (QoS) constraints, and interference-limited environments. As such, network scheduling - the problem of deciding which nodes can access the channel at any given time - becomes a critical, yet highly complex, decision-making challenge.

Traditional scheduling algorithms in wireless networks, such as Max-Weight Scheduling (MWS), focus on maximizing long-term throughput for packets without strict delay constraints. MWS is throughput-optimal, achieving any feasible throughput vector, while simpler policies like Longest-Queue-First (LQF) \cite{hu2018highest} or distributed CSMA (\cite{zhang2025distributed, huang2022distributed}) reduce computational overhead but may only attain a fraction of the throughput region. However, emerging applications require deadline-aware scheduling, which introduces a discontinuity in packet utility, making scheduling more challenging than traditional frameworks. Existing approaches to deadline-aware scheduling often rely on restrictive assumptions. Frame-based traffic models \cite{li2013optimal}, where time is divided into frames, and arrivals/deadlines are known a priori. Here, the optimal solution per frame is a Max-Weight schedule with weights based on deficit counters (tracking unmet delivery ratios). However, this approach is non-causal unless traffic is periodic and synchronized. Partial generalizations in \cite{deng2017timely} lack performance guarantees. The greedy policies like Largest-Deficit-First (LDF) \cite{kang2013performance} extend LQF to real-time scheduling, achieving an efficiency ratio (fraction of the real-time throughput region guaranteed) of at least $\frac{1}{\beta +1}$, under i.i.d. arrivals and no fading, where $\beta$  is the network’s interference degree \cite{tsanikidis2022randomized}. There have been works (\cite{wan2023scheduling, tsanikidis2021power}) that introduce randomized algorithms (AMIX-ND, AMIX-MS) for improving efficiency ratios with relaxed traffic assumptions, but face implementation barriers in large-scale networks.

Furthermore, data-driven methods such as Reinforcement Learning (RL) (\cite{yang2022joint, fawaz2021reinforcement, dutta2025cooperative}) have shown promise, but often fail to guarantee feasibility under domain-specific constraints, such as interference or packet deadlines. Furthermore, purely model-free agents tend to lack interpretability and require significant training data, making them unsuitable for rapidly changing network conditions.


With the goal of ameliorating the above limitations and inspired by cognitive architectures in neuroscience and artificial intelligence, we propose a novel hybrid scheduling framework, built using the concept of \textit{Dual Mind World Model (DMWM)} \cite{wang2025dmwm}. The proposed approach combines reactive heuristics with symbolic planning within a network digital twin platform \cite{raza2024exploiting}. The developed DMWM-inspired NDT maintains a \textit{fast reactive module} (the ``Fast Mind'') for heuristic-driven decision making, and a \textit{deliberative symbolic module} (the ``Slow Mind'') that uses internal simulation (imagination) to predict the consequences of actions over a finite horizon. Feasibility is enforced using a \textit{Informed Constraint Navigation (ICN)} layer, which acts as a symbolic constraint checker for interference, deadlines, and queue availability.

DMWM operates in a dual-phase loop: \textit{Fast Mind} triggers lightweight scheduling when constraints are simple or imagination fails; and \textit{Slow Mind} evaluates all feasible scheduling candidates using forward rollouts within the digital twin environment, selecting the schedule that optimizes long-term throughput or delay. Unlike black-box neural planners, DMWM offers \textit{interpretable planning}, \textit{constraint compliance}, and \textit{real-time responsiveness}. Its modular design allows for direct integration within real or emulated NDTs.


Specifically, this paper has the following key contributions:

\begin{enumerate}
	\item We introduce a hybrid cognitive scheduling model that integrates a symbolic planner with a reactive fallback mechanism, enabling both adaptability and low-latency execution.
	
	\item We design an efficient symbolic state-transition function that allows the agent to simulate future queue states, providing long-term foresight without neural dynamics modeling.
	
	\item We formulate a logic-based constraint filtering mechanism that enforces per-node deadlines, interference avoidance, and queue-awareness before planning.
	
	\item We construct a simulation-driven digital twin environment that models bursty traffic, heterogeneous nodes, and scheduling constraints, enabling comprehensive performance benchmarking.
	
	\item We benchmark DMWM against classical and learning-based scheduling baselines under multiple realistic scenarios, demonstrating its superiority in delay management, throughput, and feasibility adherence.
\end{enumerate}


\section{Related Work}
\label{secii}

The challenge of scheduling in wireless networks under deadline, interference, and delay constraints has been previously studied, yet many existing approaches fall short in developing solutions for dynamic, heterogeneous networks.

\textbf{Deadline-Constrained Scheduling:} Traditional scheduling policies such as Max-Weight Scheduling (MWS) offer throughput-optimal performance but are agnostic to real-time constraints~\cite{li2013optimal}. Extensions like Largest-Deficit-First (LDF)~\cite{kang2013performance} and its analytical generalization via the Real-Time Local Pooling Factor (R-LPF)~\cite{tsanikidis2021power} provide bounded efficiency ratios under specific traffic assumptions. Recent works have explored randomized scheduling~\cite{tsanikidis2022randomized}, offering improved delivery ratios, particularly under stochastic arrivals and fading channels. However, these methods often assume frame-based traffic or rely on offline knowledge of traffic statistics, limiting their adaptability in dynamic settings.

\textbf{Learning-Based Scheduling:} Reinforcement learning (RL) has gained traction as a means to learn scheduling policies from interactions with the network environment. For instance, cooperative RL schemes~\cite{dutta2025cooperative} enable decentralized scheduling in energy-harvesting IoT networks, while others apply RL in full-duplex communication environments to bypass CSI estimation~\cite{fawaz2021reinforcement}. More recently, offline RL methods~\cite{wan2023scheduling} have been proposed to train deadline-aware policies using behavior cloning and actor-critic learning. Despite their promise, most RL-based schedulers are model-free, requiring large training samples and offering limited interpretability, which poses challenges in real-time and safety-critical applications.

\textbf{Hybrid and Graph-Based Approaches:} Graph convolutional network (GCN)-based DRL models have been used for joint optimization of time-sensitive and best-effort traffic~\cite{yang2022joint}, improving generalization and convergence speed. Similarly, fast-mixing CSMA variants~\cite{li2013optimal} and distributed CSMA protocols with centralized coordination~\cite{huang2022distributed} have shown improved scalability. However, such approaches still face limitations in integrating foresight and constraint reasoning within the scheduling loop.

\textbf{Symbolic and Model-Based Methods:} Foundational work by Deng et al.~\cite{deng2017timely} characterized the delay constrained capacity region for general traffic patterns via MDPs, laying the groundwork for model-based scheduling. While their solution achieves theoretical optimality, the complexity of infinite-horizon MDPs restricts practical deployment. Input-queued switch scheduling strategies like HRF and CHRF~\cite{hu2018highest} showcase the potential of rank-based arbitration, but require specialized queue structures and are difficult to generalize for wireless networks.

\textbf{Positioning of This Work:} Compared to the aforementioned techniques, our proposed \emph{Dual Mind World Model (DMWM)} framework takes a hybrid cognitive approach, combining symbolic planning with reactive heuristics in a simulation-driven \emph{Network Digital Twin (NDT)} environment. Unlike purely model-free RL agents, DMWM performs forward rollouts over a logic-driven world model, enabling constraint-aware scheduling decisions with minimal training. Our design bridges the interpretability and efficiency of symbolic schedulers with the adaptability of learning agents, yielding superior performance under bursty, interference-limited, and deadline-sensitive conditions.

\section{System Model}
We consider a slotted-time wireless network consisting of $N$ sensor nodes, denoted by the set $\mathcal{N} = \{1, 2, \dots, N\}$. Each node maintains a local buffer (queue) and competes for channel access through centralized scheduling. The system operates under constraints typical of resource-constrained IoT networks, including interference, buffer overflows, and deadline-sensitive traffic.

\subsection{Physical Network Model}
At each time slot $t$, the queue length at node $i \in \mathcal{N}$ is denoted by $q_i(t) \in \mathbb{N}_0$. Packets arrive at node $i$ according to a Poisson process with rate $\lambda_i(t)$, which may vary with time to model bursty or periodic behavior. Nodes transmit based on a schedule $\mathcal{S}(t) \subseteq \mathcal{N}$, where $|\mathcal{S}(t)| \leq K$ is the maximum number of simultaneous transmissions per slot, dictated by the number of time-frequency resources.

We define a set of \textit{conflict pairs} $\mathcal{C} \subset \mathcal{N} \times \mathcal{N}$ to encode interference constraints. A valid schedule must satisfy:
\begin{equation}
	\forall (i,j) \in \mathcal{C},\quad i \in \mathcal{S}(t) \Rightarrow j \notin \mathcal{S}(t)
\end{equation}

Packets experience delays, and some nodes have deadline constraints $D_i$. A packet arriving at time $t_a$ must be transmitted by $t_a + D_i$, otherwise it incurs a deadline violation. The goal is to maximize long-term throughput while minimizing queue backlog, average delay, and deadline violations.

\subsection{Network Digital Twin Environment}
To enable planning and evaluation, we construct a \textit{Network Digital Twin} that serves as a virtual mirror of the physical network. It models the dynamics of:
\begin{itemize}
	\item \textbf{Traffic}: time-varying Poisson arrivals with configurable burstiness.
	\item \textbf{Queuing}: bounded FIFO queues with maximum buffer size $B$.
	\item \textbf{Delays and Deadlines}: per-packet age tracking for deadline evaluation.
	\item \textbf{Constraints}: interference (via $\mathcal{C}$) and transmission limits.
\end{itemize}

The digital twin supports \textit{symbolic imagination}—a simulated forward pass from a given state under hypothetical actions—without interacting with the real network. This facilitates the implementation of foresight-driven decision-making under the DMWM framework.

\subsection{Scheduling Actions and Rewards}
At each slot $t$, the scheduler selects a feasible action $\mathcal{S}(t)$ that:
\begin{enumerate}
	\item Satisfies all logic constraints (non-empty queues, conflict avoidance, deadlines).
	\item Maximizes a reward signal based on queue state and prediction of future congestion.
\end{enumerate}

Each successfully transmitted packet yields unit reward. The total system reward at time $t$ is:
\begin{equation}
	r(t) = \sum_{i \in \mathcal{S}(t)} \mathbb{I}[q_i(t) > 0]
\end{equation}

To support planning, the digital twin is used to simulate $H$-step rollouts from the current state:
\begin{equation}
	R(\mathcal{S}) = \sum_{\tau=0}^{H-1} \sum_{i \in \mathcal{N}} \mathbb{I}[q_i^{(\tau)} > 0]
\end{equation}
where $q_i^{(\tau)}$ represents the simulated queue length of node $i$ at future step $\tau$.

\subsection{Dual-Mind Decision Architecture}
The scheduler operates under a \textit{Dual Mind World Model (DMWM)}:

\begin{itemize}
	\item \textbf{Slow Mind (Symbolic Planner)}: Uses the digital twin to perform multi-step simulation over all valid scheduling candidates. Selects the schedule that maximizes predicted reward $R(\mathcal{S})$.
	
	\item \textbf{Fast Mind (Reactive Heuristic)}: Falls back to simple rules (e.g., LQF) when imagination is inconclusive, infeasible, or too expensive to compute in time-constrained environments.
\end{itemize}

Constraint checking is implemented using an Informed Constrained Navigation (ICN) module that filters actions based on feasibility, ensuring compliance with protocol logic.

\begin{figure}[!t]
	\centering
	\includegraphics[width=0.85\linewidth]{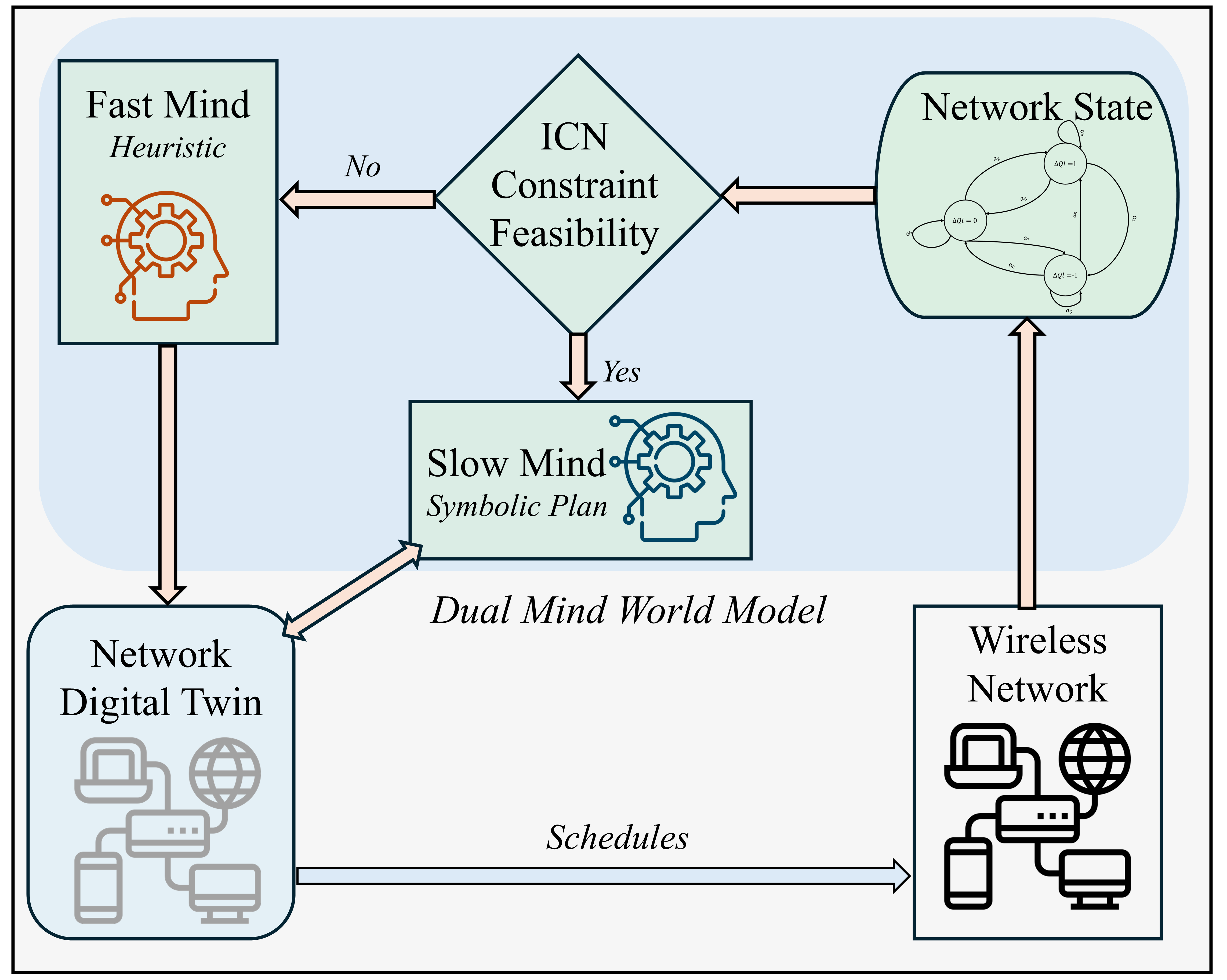}%
	\caption{System Model}
	\label{fig:sysmodel}
\end{figure}

\section{Dual Mind World Model Driven Scheduling}
The proposed scheduling architecture is based on a cognitively inspired framework that mimics dual-process decision-making in intelligent agents. This dual-mind architecture, as shown in Figure \ref{fig:sysmodel} allows the scheduler to balance between \textit{speed} and \textit{foresight}, switching modes dynamically based on current feasibility, complexity, and urgency.

\subsection{Fast Mind: Reactive Scheduling}
The fast mind selects a schedule $\mathcal{S}_\text{fast}(t)$ using a heuristic based on current queue lengths:
\begin{equation}
	\mathcal{S}_\text{fast}(t) = \arg\max_{\mathcal{S} \subset \mathcal{N}} \sum_{i \in \mathcal{S}} q_i(t), \quad |\mathcal{S}| = K
\end{equation}
This policy is computationally lightweight and ensures responsiveness under high-load or emergency conditions but lacks long-term optimality.

\subsection{Slow Mind: Symbolic Rollout Planning}
The slow mind considers a set of feasible candidate schedules:
\begin{equation}
	\mathcal{F}(t) = \left\{ \mathcal{S}_c \subset \mathcal{N} : |\mathcal{S}_c| = K,\ \text{ICN}(\mathcal{S}_c, q(t), t) = \texttt{True} \right\}
\end{equation}
where ICN is a constraint-checking function that verifies:
\begin{itemize}
	\item Non-empty queues: $q_i(t) > 0$ for all $i \in \mathcal{S}_c$
	\item Interference constraints: $(i, j) \notin \mathcal{C}$ for all $i, j \in \mathcal{S}_c$
	\item Deadline feasibility: if $D_i$ exists, schedule packets within allowable time
\end{itemize}

For each feasible schedule $\mathcal{S}_c \in \mathcal{F}(t)$, the symbolic world model simulates a rollout of horizon $H$:
\begin{equation}
	q_i^{(\tau+1)} = 
	\begin{cases}
		\max(q_i^{(\tau)} - 1, 0), & \text{if } i \in \mathcal{S}_c \\
		q_i^{(\tau)}, & \text{otherwise}
	\end{cases}
\end{equation}
starting from $q_i^{(0)} = q_i(t)$. The cumulative symbolic reward over the horizon is defined as:
\begin{equation}
	R(\mathcal{S}_c) = \sum_{\tau=0}^{H-1} \sum_{i \in \mathcal{N}} \mathbb{I}[q_i^{(\tau)} > 0]
\end{equation}

The slow mind then selects the rollout-optimal schedule:
\begin{equation}
	\mathcal{S}_\text{slow}(t) = \arg\max_{\mathcal{S}_c \in \mathcal{F}(t)} R(\mathcal{S}_c)
\end{equation}

\subsection{Combined Decision Rule}
The final scheduling decision at time $t$ is based on the availability of feasible plans:
\begin{equation}
	\mathcal{S}(t) = 
	\begin{cases}
		\mathcal{S}_\text{slow}(t), & \text{if } \mathcal{F}(t) \neq \emptyset \\
		\mathcal{S}_\text{fast}(t), & \text{otherwise}
	\end{cases}
\end{equation}

This rule ensures the best available action is chosen—either from predicted symbolic outcomes (slow mind) or from fast heuristics when symbolic planning fails or is too expensive.

\subsection{Complexity and Feasibility}
The complexity of the slow mind’s planning step is $O\left( \binom{N}{K} \cdot H \cdot N \right)$, stemming from simulating $H$-step queue evolution for each of the $\binom{N}{K}$ candidate schedules. In practice, the ICN filter significantly prunes the search space by rejecting infeasible or illogical candidates early.

\subsection{Integration with the Digital Twin}
All symbolic simulations are performed within the \textit{Network Digital Twin}, which replicates the queuing, interference, and traffic dynamics of the physical system. By using the twin for internal rollouts, the slow mind achieves foresight without affecting real-time operation. This separation enables predictive evaluation, safety checking, and explainability within a unified decision framework. The detailed working of the proposed scheduling framework is given in Algorithm \ref{alg:dmwm}, with the ICN module in Algorithm \ref{alg:icn}.

\begin{algorithm}[t]
	\caption{DMWM-Enabled Scheduling}
	\label{alg:dmwm}
	\begin{algorithmic}[1]
		\REQUIRE Current queue state $q(t)$, deadline vector $D = \{D_1, \ldots, D_N\}$, conflict graph $\mathcal{C}$, number of slots $K$, planning horizon $H$
		\ENSURE Selected schedule $S(t) \subseteq \{1, \ldots, N\}$ with $|S(t)| = K$
		\STATE $\mathcal{F} \leftarrow \emptyset$ \COMMENT{Feasible schedule set}
		\FORALL{$S \subseteq \{1, \ldots, N\}$ with $|S| = K$}
		\IF{\textsc{ICN\_Check}($S$, $q(t)$, $D$, $\mathcal{C}$, $t$)}
		\STATE $\mathcal{F} \leftarrow \mathcal{F} \cup \{S\}$
		\ENDIF
		\ENDFOR
		
		\IF{$\mathcal{F} \neq \emptyset$}
		\STATE $R^\star \leftarrow -\infty$, $S^\star \leftarrow \emptyset$
		\FORALL{$S \in \mathcal{F}$}
		\STATE $q^{(0)} \leftarrow q(t)$, $R \leftarrow 0$
		\FOR{$\tau = 0$ to $H - 1$}
		\FORALL{$i \in S$}
		\STATE $q_i^{(\tau+1)} \leftarrow \max(q_i^{(\tau)} - 1, 0)$
		\ENDFOR
		\STATE $R \leftarrow R + \sum_{i=1}^{N} \mathbb{I}[q_i^{(\tau)} > 0]$
		\ENDFOR
		\IF{$R > R^\star$}
		\STATE $R^\star \leftarrow R$, $S^\star \leftarrow S$
		\ENDIF
		\ENDFOR
		\RETURN $S^\star$
		\ELSE
		\FOR{$i = 1$ to $N$}
		\STATE $u_i \leftarrow q_i(t) \cdot (2 \text{ if } D_i < \infty \text{ else } 1)$
		\ENDFOR
		\STATE $S_{\text{fallback}} \leftarrow$ top-$K$ indices by $u_i$
		\RETURN $S_{\text{fallback}}$
		\ENDIF
	\end{algorithmic}
\end{algorithm}

\begin{algorithm}[t]
	\caption{\textsc{ICN\_Check}$(S, q(t), D, \mathcal{C}, t)$}
	\label{alg:icn}
	\begin{algorithmic}[1]
		\FORALL{$i \in S$}
		\IF{$q_i(t) = 0$}
		\RETURN \textbf{False}
		\ENDIF
		\IF{$D_i < \infty$ and oldest packet age $> D_i$}
		\RETURN \textbf{False}
		\ENDIF
		\ENDFOR
		\FORALL{$(i, j) \in S \times S$}
		\IF{$(i, j) \in \mathcal{C}$}
		\RETURN \textbf{False}
		\ENDIF
		\ENDFOR
		\RETURN \textbf{True}
	\end{algorithmic}
\end{algorithm}

\begin{table}[htbp]
	\caption{Baseline Simulation Hyperparameters}
	\centering
	\begin{tabular}{|l|c|}
		\hline
		\textbf{Parameter} & \textbf{Value} \\
		\hline
		Number of Nodes ($N$) & 5 \\
		Max Scheduled Nodes ($K$) & 3 \\
		Queue Capacity ($B$) & 50 \\
		Time Steps per Run & 200 \\
		Poisson Arrival Base Range ($\lambda_i^{\text{base}}$) & [0.5, 1.0] \\
		Imagination Horizon ($H$) & 3 \\
		Q-Learning $\alpha$ (Learning Rate) & 0.1 \\
		Q-Learning $\gamma$ (Discount Factor) & 0.95 \\
		Q-Learning $\epsilon$ (Exploration Rate) & 0.2 \\
		\hline
	\end{tabular}
	\label{tab:params}
\end{table}

\section{Experimental Setup}
\label{exp_setup}

To evaluate the proposed Dual Mind World Model (DMWM) scheduler, we implement a simulation-driven digital twin of a time-slotted wireless network. The digital twin captures queue dynamics, stochastic arrivals, interference constraints, and per-node deadlines, in addition to forwarding the scheduling policies learnt by the DMWM scheduler. All experiments are carried out across multiple network scenarios with diverse constraints, compared against baseline policies.

\subsection{Network Simulation Environment}

We simulate a network of $N$ nodes, each maintaining a bounded FIFO queue of size $B=50$. Time is slotted, and up to $K$ nodes can be scheduled in each slot. Traffic arrivals at each node follow a time-varying Poisson process with dynamic arrival rates $\lambda_i(t)$. The traffic profile includes both smooth and bursty patterns to mimic real-world variability:
\[
\lambda_i(t) = \lambda_i^{\text{base}} \cdot \left(1 + 0.75 \cdot \sin\left( \frac{2\pi t}{50} \right) \right) + \delta_i(t)
\]
where $\delta_i(t)$ introduces occasional burst spikes on selected nodes.

Scheduling actions are constrained by a conflict graph $\mathcal{C}$ (interference model) and optional per-node packet deadlines $D_i$.

\subsection{Evaluation Scenarios}

We evaluate the policies in four distinct scenarios:

\begin{itemize}
	\item \textbf{Default}: Moderate interference (pairs of conflicting nodes) and mixed deadline constraints on alternate nodes.
	\item \textbf{Bursty Traffic}: High variance in arrival rates with no deadlines or conflicts.
	\item \textbf{Deadline-Sensitive}: Alternating nodes with short (5) and long (15) deadlines, plus pairwise interference.
	\item \textbf{Interference-Constrained}: Circular conflict graph without deadlines.
\end{itemize}

\subsection{Baseline Scheduling Policies}

The proposed DMWM scheduler is compared against the following baselines:

\begin{itemize}
	\item \textbf{Random}: Selects $K$ nodes uniformly at random.
	\item \textbf{Longest Queue First (LQF)}: Prioritizes nodes with highest queue length.
	\item \textbf{Deadline-Priority}: Assigns urgency weights based on packet deadlines.
	\item \textbf{Fair Round Robin}: Ensures fairness over time using a memory-based rotation.
	\item \textbf{RL-Agent (Q-Learning)}: Online learning-based scheduler with $\epsilon$-greedy exploration, updated via Q-table.
\end{itemize}

The DMWM agent uses a symbolic forward model (imagination) with a planning horizon of $H=3$ to evaluate candidate schedules filtered through the ICN constraint layer.

\begin{figure}[!h]
	\centering
	\includegraphics[width=0.8\linewidth, height=1.1\linewidth]{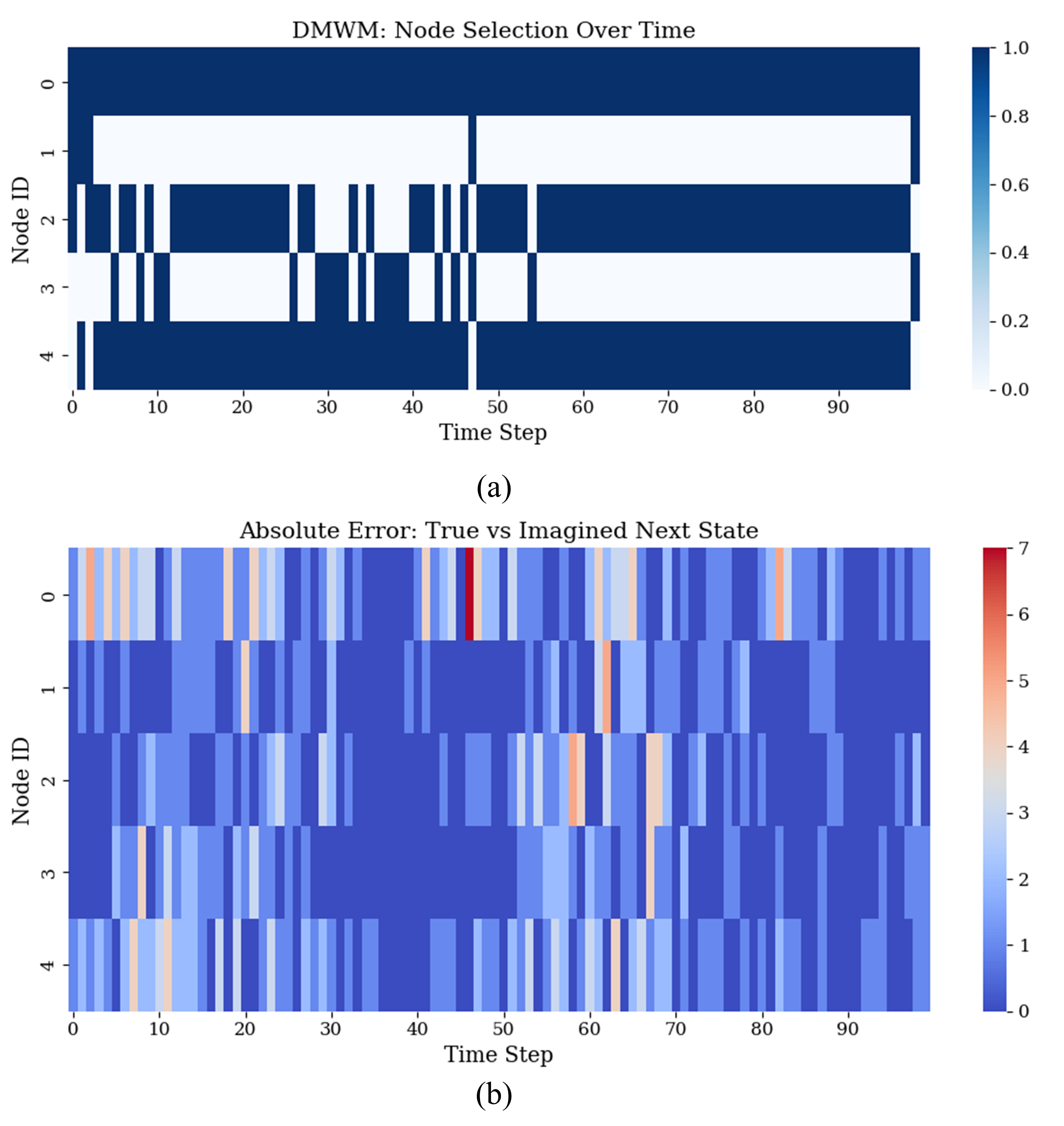}%
	\caption{DMWM Scheduler Behavior and Model Accuracy: This plot illustrates the node scheduling behavior of the DMWM policy, where each row corresponds to a node, and each column represents a time step. A blue cell (in (a)) indicates that the corresponding node was selected for transmission at that time. Subplot (b) visualizes the absolute difference between the actual queue state transitions and those predicted by the symbolic world model used in the Slow Mind's internal simulations.}
	\label{fig:dmwm_visual}
\end{figure}

\subsection{Hyperparameters and Performance Metrics}

All policies are tested over $30$ independent simulation runs, each lasting $200$ time steps. The baseline parameters used in the simulation environment are summarized in Table~\ref{tab:params}. We evaluate each policy using the following metrics: average throughput, queue length, delay, and deadline violations. Each metric is reported with its mean and standard deviation over multiple simulation runs.

\section{Results and Analysis}

To assess the effectiveness of the proposed DMWM-inspired scheduling framework, we conduct extensive simulations across diverse network conditions that mimic real-world challenges in IoT and wireless access systems. As explained earlier in section \ref{exp_setup}, we compare the performance of DMWM against traditional baselines and a reinforcement learning (RL) agent under four distinct traffic and constraint scenarios: default, bursty, deadline-sensitive, and interference-limited. First, to gain insight into the internal dynamics of DMWM, we analyze two behavioral visualizations shown in Figure \ref{fig:dmwm_visual}. The heatmap shown in Figure \ref{fig:dmwm_visual} (a) illustrates the node scheduling pattern selected by DMWM over time. The schedule adapts dynamically to queue states and constraint violations. The diversity and structure in activation patterns demonstrate the interplay between the reactive (Fast Mind) and symbolic (Slow Mind) modules. A comparison between imagined and actual next states (Figure \ref{fig:dmwm_visual} (a)) reveals low absolute prediction errors across nodes, confirming the reliability of the internal model used for forward simulations. This validates the ability of the digital twin to anticipate the dynamics of the queues over the planning horizon.

\begin{figure*}[!h]
	\centering
	\includegraphics[width=0.8\linewidth]{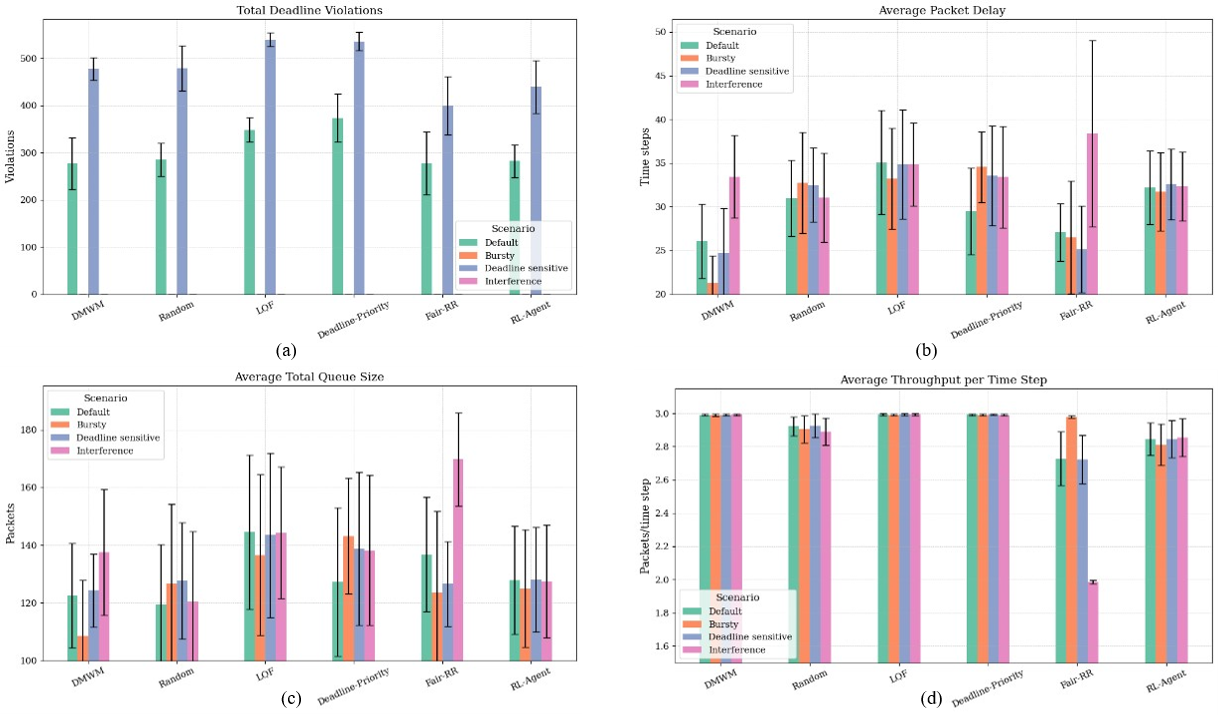}%
	\caption{This figure compares six scheduling policies—DMWM, Random, LQF, Deadline-Priority, Fair Round-Robin, and a Q-learning agent—across four scenarios: Default, Bursty, Deadline-Sensitive, and Interference-Constrained. Results show mean performance over 30 runs with standard deviation bars. DMWM attains the best or near-best throughput while maintaining low delay and queue occupancy, achieving efficient utilization without violating interference or deadline constraints.}
	\label{fig:performance_subplots}
\end{figure*}

Figure \ref{fig:performance_subplots} presents comparative results of all policies across four scenarios, averaged over multiple runs. The following key findings are observed. Firt, DMWM-enabled scheduling consistently achieves the highest or near-highest average throughput across all scenarios. Notably, it outperforms both LQF and the RL-agent under bursty and deadline-sensitive loads, due to its planning-based imagination module that anticipates congestion and avoids conflicts. Second, in terms of queue occupancy, DMWM maintains lower average queue sizes compared to reactive policies, indicating more effective congestion mitigation. Particularly in interference-heavy scenarios, its symbolic constraint module allows it to navigate contention zones more efficiently than LQF and Random. Finally, the most significant improvements from DMWM are observed in latency-sensitive metrics. Under the deadline-sensitive scenario, DMWM reduces the average delay and number of deadline violations substantially relative to the RL-agent and heuristics. This is attributed to its forward rollout mechanism, which explicitly evaluates the impact of delay during scheduling. To be noted that DMWM not only performs well on average but also shows lower variance across trials, indicating robustness under dynamic network conditions.

\section{Conclusion}
\label{sec:conclusion}

In this paper, we proposed a novel Dual Mind World Model (DMWM) approach for network scheduling, designed to operate within a Network Digital Twin (NDT) simulation environment. The DMWM framework integrates symbolic planning and lightweight model-based prediction, offering a balance between adaptability and interpretability. By leveraging imagined state transitions over a short planning horizon, DMWM allows the scheduler to anticipate and mitigate network congestion, interference, and deadline violations without requiring deep function approximators or extensive retraining. We evaluated the proposed scheduler across multiple simulated network scenarios characterized by bursty traffic, deadline constraints, and interference. Comparative results against classic baselines demonstrated DMWM’s robustness, particularly in non-stationary or constrained environments. In addition to superior performance in key metrics like throughput, delay, and deadline adherence, DMWM offers a transparent decision-making process through interpretable rollout mechanisms. This makes it suited for critical or resource-constrained systems requirung accountability and insight.

	\ifCLASSOPTIONcaptionsoff
	\newpage
	\fi

	\bibliographystyle{IEEEtran}
	\bibliography{bibtex/Reference}

\begin{thebibliography}{10}
\providecommand{\url}[1]{#1}
\csname url@samestyle\endcsname
\providecommand{\newblock}{\relax}
\providecommand{\bibinfo}[2]{#2}
\providecommand{\BIBentrySTDinterwordspacing}{\spaceskip=0pt\relax}
\providecommand{\BIBentryALTinterwordstretchfactor}{4}
\providecommand{\BIBentryALTinterwordspacing}{\spaceskip=\fontdimen2\font plus
\BIBentryALTinterwordstretchfactor\fontdimen3\font minus
  \fontdimen4\font\relax}
\providecommand{\BIBforeignlanguage}[2]{{%
\expandafter\ifx\csname l@#1\endcsname\relax
\typeout{** WARNING: IEEEtran.bst: No hyphenation pattern has been}%
\typeout{** loaded for the language `#1'. Using the pattern for}%
\typeout{** the default language instead.}%
\else
\language=\csname l@#1\endcsname
\fi
#2}}
\providecommand{\BIBdecl}{\relax}
\BIBdecl

\bibitem{hu2018highest}
B.~Hu, F.~Fan, K.~L. Yeung, and S.~Jamin, ``Highest rank first: A new class of
  single-iteration scheduling algorithms for input-queued switches,''
  \emph{IEEE Access}, vol.~6, pp. 11\,046--11\,062, 2018.

\bibitem{zhang2025distributed}
Z.~Zhang, S.~Atapattu, Y.~Wang, and M.~Di~Renzo, ``Distributed mac for
  ris-assisted multiuser networks: Csma/ca protocol design and statistical
  optimization,'' \emph{IEEE Transactions on Mobile Computing}, 2025.

\bibitem{huang2022distributed}
C.~Huang and X.~Wang, ``Distributed scheduling with centralized coordination
  for scalable wireless mesh networking,'' \emph{IEEE/ACM Transactions on
  Networking}, vol.~31, no.~1, pp. 436--451, 2022.

\bibitem{li2013optimal}
B.~Li and A.~Eryilmaz, ``Optimal distributed scheduling under time-varying
  conditions: A fast-csma algorithm with applications,'' \emph{IEEE
  Transactions on Wireless Communications}, vol.~12, no.~7, 2013.

\bibitem{deng2017timely}
L.~Deng, C.-C. Wang, M.~Chen, and S.~Zhao, ``Timely wireless flows with general
  traffic patterns: Capacity region and scheduling algorithms,'' \emph{IEEE/ACM
  Transactions on Networking}, vol.~25, no.~6, 2017.

\bibitem{kang2013performance}
X.~Kang, W.~Wang, J.~J. Jaramillo, and L.~Ying, ``On the performance of
  largest-deficit-first for scheduling real-time traffic in wireless
  networks,'' in \emph{ACM international symposium on Mobile ad hoc networking
  and computing}, 2013, pp. 99--108.

\bibitem{tsanikidis2022randomized}
C.~Tsanikidis and J.~Ghaderi, ``Randomized scheduling of real-time traffic in
  wireless networks over fading channels,'' \emph{IEEE/ACM Transactions on
  Networking}, vol.~31, no.~4, pp. 1688--1701, 2022.

\bibitem{wan2023scheduling}
J.~Wan, S.~Lin, Z.~Zhang, J.~Zhang, and T.~Zhang, ``Scheduling real-time
  wireless traffic: A network-aided offline reinforcement learning approach,''
  \emph{IEEE internet of things journal}, vol.~10, no.~24, 2023.

\bibitem{tsanikidis2021power}
C.~Tsanikidis and J.~Ghaderi, ``On the power of randomization for scheduling
  real-time traffic in wireless networks,'' \emph{IEEE/ACM Transactions on
  Networking}, vol.~29, no.~4, pp. 1703--1716, 2021.

\bibitem{yang2022joint}
L.~Yang, Y.~Wei, F.~R. Yu, and Z.~Han, ``Joint routing and scheduling
  optimization in time-sensitive networks using
  graph-convolutional-network-based deep reinforcement learning,'' \emph{IEEE
  Internet of Things Journal}, vol.~9, no.~23, pp. 23\,981--23\,994, 2022.

\bibitem{fawaz2021reinforcement}
H.~Fawaz, M.~El~Helou, S.~Lahoud, and K.~Khawam, ``A reinforcement learning
  approach to queue-aware scheduling in full-duplex wireless networks,''
  \emph{Computer Networks}, vol. 189, p. 107893, 2021.

\bibitem{dutta2025cooperative}
H.~Dutta, A.~K. Bhuyan, and S.~Biswas, ``Cooperative reinforcement learning for
  energy management in multi-hop networks with energy harvesting,'' \emph{IEEE
  Transactions on Green Communications and Networking}, 2025.

\bibitem{wang2025dmwm}
L.~Wang, R.~Shelim, W.~Saad, and N.~Ramakrishnan, ``Dmwm: Dual-mind world model
  with long-term imagination,'' \emph{arXiv preprint arXiv:2502.07591}, 2025.

\bibitem{raza2024exploiting}
S.~M. Raza, R.~Minerva, N.~Crespi, M.~Alvi, M.~Herath, and H.~Dutta,
  ``Exploiting the efficient data modeling in network digital twin to empower
  edge-cloud continuum,'' in \emph{2024 20th International Conference on
  Network and Service Management (CNSM)}.\hskip 1em plus 0.5em minus
  0.4em\relax IEEE, 2024, pp. 1--4.

\end{thebibliography}

\end{document}